\documentclass[
reprint,
superscriptaddress,
amsmath,amssymb,
aps,
prb,
floatfix
]{revtex4-2}

\usepackage{graphicx}
\usepackage{dcolumn}
\usepackage{bm}
\usepackage[colorlinks,citecolor=magenta]{hyperref}
\usepackage[T1]{fontenc}

\begin{document}

\title{Couplings between Photons and Ensemble of Emitters in a High-$Q$ Cavity at Mesoscopic Excitation Levels}
\thanks{© 2025 IEEE.  Personal use of this material is permitted.  Permission from IEEE must be obtained for all other uses, in any current or future media, including reprinting/republishing this material for advertising or promotional purposes, creating new collective works, for resale or redistribution to servers or lists, or reuse of any copyrighted component of this work in other works.}

\author{Chenjiang Qian}
\email{chenjiang.qian@iphy.ac.cn}
\affiliation{Walter Schottky Institut and TUM School of Natural Sciences, Technische Universit{\" a}t M{\" u}nchen, Am Coulombwall 4, 85748 Garching, Germany}
\affiliation{Beijing National Laboratory for Condensed Matter Physics, Institute of Physics, Chinese Academy of Sciences, Beijing 100190, China}
\affiliation{School of Physical Sciences, University of Chinese Academy of Sciences, Beijing 100049, China}
\author{Viviana Villafañe}
\affiliation{Walter Schottky Institut and TUM School of Natural Sciences, Technische Universit{\" a}t M{\" u}nchen, Am Coulombwall 4, 85748 Garching, Germany}
\author{Pedro Soubelet}
\affiliation{Walter Schottky Institut and TUM School of Natural Sciences, Technische Universit{\" a}t M{\" u}nchen, Am Coulombwall 4, 85748 Garching, Germany}
\author{Andreas V. Stier}
\affiliation{Walter Schottky Institut and TUM School of Natural Sciences, Technische Universit{\" a}t M{\" u}nchen, Am Coulombwall 4, 85748 Garching, Germany}
\author{Jonathan J. Finley}
\email{finley@wsi.tum.de}
\affiliation{Walter Schottky Institut and TUM School of Natural Sciences, Technische Universit{\" a}t M{\" u}nchen, Am Coulombwall 4, 85748 Garching, Germany}

\begin{abstract}
  We investigate the coupling between an ensemble of individual emitters and multiple photons in a high-$Q$ cavity at the mesoscopic excitation level.
  The master equation theory is used to calculate the emission spectrum of the cavity QED system.
  The increasing excitation level not only pumps the system to the high-energy multi-emitter-photon states, but also introduces the pump-induced dephasing that suppresses the coherent energy exchange (coupling) between photons and emitters.
  When the emitter lifetime exceeds a threshold, we observe the mesoscopic excitation level i.e., the system is pumped to high energy states whilst the coherent couplings between these states are not yet suppressed.
  The mesoscopic excitation enables the couplings between multi-emitter-photon states, and thereby, paves the way to building quantum photonic devices based on multiple photons and nonlinear effects.
\end{abstract}

\maketitle

\section{Introduction}

The cavity QED system consisting of embedded emitters and confined photons provides a quantum interface between light and matter degrees of freedom \cite{doi:10.1126/science.1078446,Walther_2006}.
Such nanosystem is the basic building block of the processing and storage of quantum information in the network \cite{Kimble2008,O'Brien2009,Ritter2012,RevModPhys.87.347,RevModPhys.87.1379} and is also widely applied in other fields including sensor, energy and biology \cite{Alivisatos2004,doi:10.1021/acs.jpclett.8b00008,Yu2021}.
In the limit of weak excitation, the system is at lowest energy states with either a single photon or an excited two-level emitter.
The coherent energy exchange (coupling) between them has been investigated and applied in quantum photonic devices \cite{Reithmaier2004,Yoshie2004,PhysRevLett.108.227402,Kim2013}.
In addition, the coupling strength between high energy states increases with the number of involved photons and emitters \cite{PhysRev.170.379,PhysRev.188.692,RevModPhys.80.517,PhysRevA.87.043817}.
Such high order coupling allows the emission and quantum information processing involving multiple photons, thereby is attracting increasing interest in recent years \cite{PhysRevLett.107.233602,Munoz2014,PhysRevLett.120.213901,PhysRevApplied.12.044065,PhysRevLett.122.087401,doi:10.1021/acs.nanolett.0c01562,Lo2021,PhysRevResearch.4.023052}.

The high order coupling requires the cavity QED system pumped to high energy multi-emitter-photon states.
However, as the excitation level increases, the incoherent pump introduces the dephasing which suppresses the coherent emitter-photon coupling \cite{PhysRevLett.103.087405,0906.1455,PhysRevB.81.033309,Nomura2010}.
At strong excitation levels, the pump-induced dephasing suppresses the coherent energy exchange, thus all emissions are merged to the bare cavity mode \cite{PhysRevLett.103.087405}.
As such, a mesoscopic excitation level, i.e., the system is pumped to high energy states while the coherent energy exchange between them is not yet suppressed, is required to observe the high order coupling.

Here we investigate the high order couplings between a high-$Q$ cavity mode and an ensemble of \textit{individual} emitters using the master equation theory \cite{Tan_1999}.
Pump dependent emission spectra with $N_X$ emitters and maximum $N_C$ photons are calculated.
When $N_{C(X)} \neq 1$, the system will be pumped to high energy states as the pump increases.
For narrow-linewidth emitters, the high order couplings at mesoscopic excitation levels are clearly observed from the nonmonotonic energy shift of emission peaks.
In contrast, for broad-linewidth emitters with the lifetime below the threshold, only the monotonic energy shift is observed, indicating the pump-induced dephasing dominates throughout.

We verify the calculation in a high-$Q$ nanobeam cavity embedded with a monolayer MoS$_2$ \cite{PhysRevLett.128.237403}.
The \textit{localized} excitons (LXs) are a good case of \textit{individual} emitters, since their wavefunctions are strongly localized at defects \cite{Klein2019,Mitterreiter2021,Trainer2022,10.1126/sciadv.adk6359}.
We record PL spectra of the cavity-LXs system.
As expected, the emission peak exhibits the nonmonotonic power-dependent energy shift that indicates the high order couplings, which is in contrast not observed in control experiments recorded from MoS$_2$ outside cavities.

\section{\label{sec2}Theory and Calculation}

\subsection{\label{sec2a}Multi-Emitter-Photon System}

\begin{figure*}
    \includegraphics[width=\linewidth]{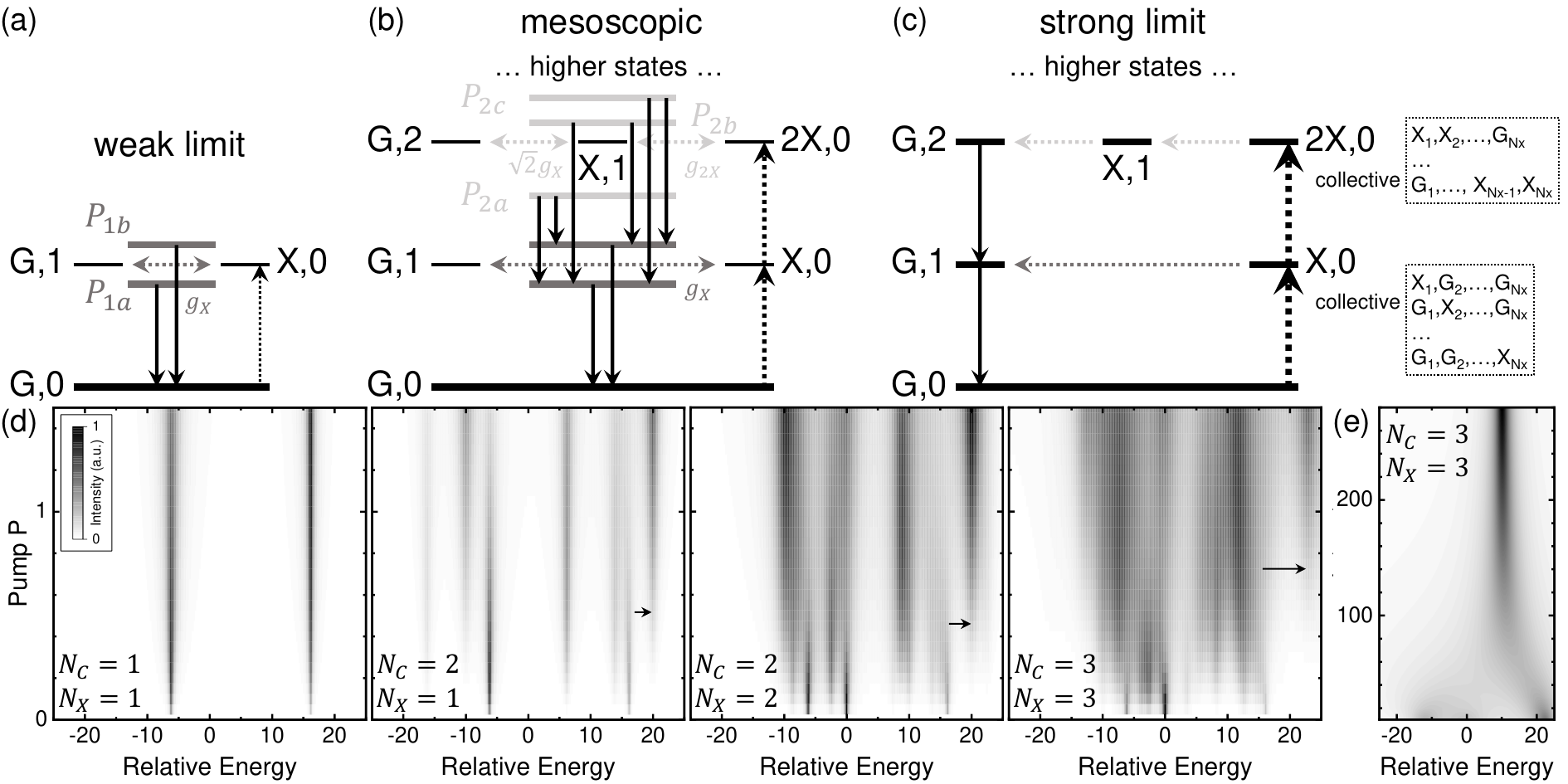}
    \caption{\label{f1}
        (a)-(c) Schematic of couplings in the cavity QED system at weak, mesoscopic and strong excitation levels, respectively.
        (d) Calculated emission spectra with a relative small pump $P=0-1.5$, in the cases of various number of emitters $N_X$ and photons $N_C$.
        Black arrows indicate that the system is pumped from low to high order couplings.
        (e) Calculated emission spectra for $N_{C(X)}=3$.
        With the strong pump $P>100$, the pump-induced dephasing suppresses all coherent energy exchanges, and the transitions degenerate to the bare cavity energy as depicted in (c).
    }
\end{figure*}

We consider a single cavity mode with energy $\omega_C$ and linewidth $\gamma_C$, coupled to an ensemble of $N_X$ \textit{individual} two-level emitters.
Each emitter has the same energy $\omega_X$, linewidth $\gamma_X$, and coupling strength $g_i$ to the cavity mode.
The state of cavity QED system consists of the state of emitters and the number of photons in the cavity.
As depicted in Fig.~\ref{f1}(a)-(c), the emitter state is collectively described by the number of excited emitters, e.g., $G$ means all emitters are at the ground level, $X$ means one emitter is excited, and $2X$ means two emitters are excited.
Following is the number of photons in the cavity mode.
$|G,0\rangle$ is thereby the ground state of the whole system.
In the framework of the rotating wave approximation, the coupling between photons and emitters is described by the Jaynes-Cummings or Tavis-Cummings model \cite{10.1109/PROC.1963.1664,PhysRev.170.379,PhysRev.188.692}.
The system Hamiltonian is given by
\begin{eqnarray}
    \label{eqH}
    \textit{H} &=& \hbar\omega_{C}a^{+}a + \sum_{i=1}^{N_X}\hbar\omega_{X}\sigma_{Xi,Xi} \nonumber \\
    &\;& + \sum_{i=1}^{N_X}\hbar g_{i}\left(\sigma_{Gi,Xi}a^{+} + \sigma_{Xi,Gi}a\right)
\end{eqnarray}
where $\sigma_{Ai,Bi}=|Ai \rangle \langle Bi|$ ($A,B \in {X,G}$) is the Dirac operator for emitter $i$ and $a^{+}$/$a$ is the ladder operators for photons.
This model is valid when the cavity mode volume is small i.e., the free spectral range
is much larger than the coupling strength \cite{PhysRevLett.123.243602,PhysRevA.105.013719}.

The simple case is the first order coupling in the weak excitation limit, as depicted in Fig.~\ref{f1}(a).
This is the focus of usual investigations.
In this case, the energy symmetrically and coherently exchanges between the $|G,1\rangle$ and $|X,0\rangle$ state.
Since the $X \rightarrow G$ transition is a collective of $N_X$ transitions, the coupling strength between these two states is $g_X=\sqrt{N_X}g_i$ \cite{PhysRev.170.379,PhysRev.188.692}.
This coupling results in the two first-order polariton eigenstates denoted by $P_{1a}$ and $P_{1b}$.
The corresponding two eigenfrequencies can be calculated by solving the Hamiltonian matrix
\begin{eqnarray}
    \label{eqH2}
    \left(
    \begin{array}{cc}
            \Omega_{C} & g_X        \\
            g_X        & \Omega_{X}
        \end{array}\right)\;
\end{eqnarray}
and results are
\begin{eqnarray}
    \Omega_\pm=\frac{\Omega_X+\Omega_C}{2}\pm\sqrt{g^2+\left(\frac{\Omega_X-\Omega_C}{2}\right)^2}\nonumber
\end{eqnarray}
where $\Omega_X=\omega_X-i\gamma_X/2$ and $\Omega_C=\omega_C-i\gamma_C/2$.
The well-known Rabi splitting at resonance $\omega_X=\omega_C$ occurs when the coupling strength is larger than the decay rate as $g_X>\left(\gamma_X-\gamma_C\right)/4$ \cite{PhysRevB.60.13276,Khitrova2006}.

In addition, the cavity QED system also supports high energy states e.g., the $|G,2\rangle$, $|X,1\rangle$ and $|2X,0\rangle$ states depicted in Fig.~\ref{f1}(b).
The coupling strength between $|G,2\rangle$ and $|X,1\rangle$ is $\sqrt{2}g_X$, where the $\sqrt{2}$ arises from the ladder operator for two photons.
The $2X \rightarrow X$ transition is a collective of $N_X\left(N_X-1\right)$ transitions, thus the coupling strength between $|X,1\rangle$ and $|2X,0\rangle$ is $g_{2X}=\sqrt{N_X\left(N_X-1\right)}g_i$ \cite{PhysRev.170.379,PhysRev.188.692}.
The couplings between these three states result in three second-order polariton eigenstates denoted by $P_{2a}$, $P_{2b}$, and $P_{2c}$.
The eigenfrequency of polaritons can be calculated by solving the Hamiltonian matrix
\begin{eqnarray}
    \label{eqH3}
    \left(
    \begin{array}{ccc}
            2\Omega_{C}   & \sqrt{2}g_{X}         & 0           \\
            \sqrt{2}g_{X} & \Omega_{C}+\Omega_{X} & g_{2X}      \\
            0             & g_{2X}                & 2\Omega_{X}
        \end{array}\right)\;.
\end{eqnarray}
As such, when the excitation is enough to pump the system to these three states, 6 emission peaks representing the transition from second-order to first-order polaritons would be observed as depicted by the downward arrows in Fig.~\ref{f1}(b).
Similar couplings can be expected between higher energy states involving more excited emitters and photons.

However, the Hamiltonian matrix in Eqs.~\ref{eqH2} and \ref{eqH3} are only valid with a relative weak excitation, which means the pump has little impact on the energy exchange between emitters and photons.
As the excitation level increases, the pump-induced dephasing will suppress the coherent energy exchange \cite{PhysRevLett.103.087405,0906.1455,PhysRevB.81.033309,Nomura2010}.
In the limit of strong excitation, emitters are approximately forced at the excited state, with only the unidirectional energy transfer from emitters to photons as depicted in Fig.~\ref{f1}(c).
In this case, the spectral emission degenerates to the bare cavity mode, which includes the transitions $|G,1\rangle\rightarrow|G,0\rangle$, $|G,2\rangle\rightarrow|G,1\rangle$, etc., no matter the number of photons and emitters in the system.

\subsection{\label{sec2b}Spectra Calculated from Master Equation}

As discussed above, the increasing excitation level has two effects: pump the system to high energy states and suppress the coherent energy exchange.
To further investigate the dominance of the two effects, we use Quantum Optics Toolbox \cite{Tan_1999} to solve the maser equation of the coupling system.
The master equation is given by
\begin{eqnarray}
    \label{master}
    \frac{d}{dt}\rho&=&-\frac{i}{\hbar}[H,\rho]+\sum_{n}{\cal L}(c_n)
\end{eqnarray}
where $\rho$ is the density matrix including all allowed states with maximum $N_C$ photons, $H$ is the Hamiltonian defined in Eq.~\ref{eqH}, and the Liouvillian superoperator
\begin{eqnarray}
    \label{decay}
    {\cal L}(c_n)=1/2\left(2c_n\rho c_n^+-\rho c_n^+c_n-c_n^+c_n\rho \right) \ \ \ \ \ \\
    c_{n}=\lbrace \sqrt{\gamma_C}a,\sqrt{\gamma_X}\sigma_{Gi,Xi},\sqrt{P}\sigma_{Xi,Gi}\rbrace, i\in[1,N_X] \nonumber
\end{eqnarray}
describes Markovian processes corresponding to the decay of photons, the decay of emitters, and the pump of emitters with rate $P$, respectively.
The pump in Eq.~\ref{decay} is incoherent thereby introduces the pump-induced dephasing \cite{PhysRevLett.103.087405,0906.1455,PhysRevB.81.033309,Nomura2010}.
For brevity, we use energy unit for all the parameters $\omega_{X\left(C\right)}$, $\gamma_{X\left(C\right)}$, and $P$ (e.g. $\omega_{X}$ is actually $\hbar\omega_{X}$) and omit the unit meV.
The emission spectrum is calculated from the steady state of Eq.~\ref{master}.

In Fig.~\ref{f1}(d) we present typical emission spectra calculated for narrow-linewidth emitters with various $N_{C(X)}$.
The parameters are set as $\omega_X=0$, $\omega_C=10$, $\gamma_X=0.25$, and $\gamma_C=0.2$.
The coupling strength of the cavity mode to a single emitter is set to $g_i=g_X/\sqrt{N_X}$, to keep a same value of the first-order coupling strength $g_X=10$ between the cases with different $N_X$.
We emphasize that the collective coupling strength e.g., $g_X$, $\sqrt{2}g_X$, and $g_{2X}$ in Eq.~\ref{eqH2} and \ref{eqH3}, is not a precondition of master equation Eq.~\ref{master} (only $g_i$ is), but is a result which can be achieved by solving the master equation.

$N_{C(X)}=1$ is the simplest case having four states of the system as $|G,0\rangle$, $|G,1\rangle$, $|X,0\rangle$, and $|X,1\rangle$.
With the relative small pump $P=0-1.5$, the coherent energy exchange between $|G,1\rangle$ and $|X,0\rangle$ results in two polariton eigenstates with the eigenfrequencies as
\begin{eqnarray}
    P_{1a}&=&0.85|X,0\rangle-0.53|G,1\rangle,\ \Omega_{P1a}=-6.2-0.12i \nonumber \\
    P_{1b}&=&0.53|X,0\rangle+0.85|G,1\rangle,\ \Omega_{P1b}=16.2-0.11i \nonumber \ .
\end{eqnarray}
Due to the degeneration of transitions $P_{1a}\rightarrow |G,0\rangle$ and $|X,1\rangle\rightarrow P_{1b}$ (energy of $-6.2$), as well as $P_{1b}\rightarrow |G,0\rangle$ and $|X,1\rangle\rightarrow P_{1a}$ (energy of $16.2$), only two emission peaks are observed in the spectra.
In contrast, with the maximum photon number $N_C=2$, the system has two additional states $|G,2\rangle$ and $|X,2\rangle$.
The coherent energy exchange between $|G,2\rangle$ and $|X,1\rangle$ results in two second-order polariton eigenstates with the eigenfrequencies
\begin{eqnarray}
    P_{2a}&=&0.82|X,1\rangle-0.58|G,2\rangle,\ \Omega_{P2a}=0.0-0.22i \nonumber \\
    P_{2b}&=&0.58|X,1\rangle+0.82|G,2\rangle,\ \Omega_{P2b}=30.0-0.21i \nonumber \ .
\end{eqnarray}
As shown in Fig.~\ref{f1}(d), with increasing pump $P$, emission peaks of the first-order transitions $P_{1a}\rightarrow |G,0\rangle$ at $-6.2$ and $P_{1b}\rightarrow |G,0\rangle$ at $16.2$ quickly suppress when $P>0.4$.
Emission peaks from the second-order transitions $P_{2a}\rightarrow P_{1a}$ at $6.2$, $P_{2a}\rightarrow P_{1b}$ at $-16.2$, and $P_{2b}\rightarrow P_{1b}$ at $13.8$ are firstly enhanced when $P<1$ and then suppressed.
The other second-order transition $P_{2b}\rightarrow P_{1a}$ at $36.2$ (out of plot range) behaviors similar to these three.
The emission peaks from highest order transitions $|X,2\rangle\rightarrow P_{2a}$ at $20$ and $|X,2\rangle\rightarrow P_{2b}$ at $-10$ are continuously enhanced in the calculation range of $P=0-1.5$.
As the pump increases, the converting of dominated emission peak is consistent with the coupling pumped from low to high orders.

When more photons and emitters are involved in the system, the coupling becomes complex, and more transitions appear in the emission spectra \cite{PhysRevLett.120.213901,PhysRevLett.133.086902} e.g., the results with $N_{C(X)}=2$ and $N_{C(X)}=3$ presented in Fig.~\ref{f1}(d).
Despite the complex details, we clearly observe the central result i.e., the increasing $P$ pumps the system from low to high order couplings, as denoted by the black arrows in Fig.~\ref{f1}(d).
In addition, although the first-order coupling strength $g_X$ is set to same, the high order coupling strength increases with $N_{C(X)}$ \cite{PhysRev.170.379,PhysRev.188.692}.
Thereby, larger energy shift between low and high order transitions are observed, e.g., by comparing the spectra with $N_{C(X)}=3$ and $N_{C(X)}=2$.
In contrast to the relative small $P$ in Fig.~\ref{f1}(d), we present the case with $N_{C(X)}=3$ but with a much larger $P$ in Fig.~\ref{f1}(e).
As $P$ approaches the strong limit $>100$, all emission peaks degenerate to the bare cavity energy at $\omega_C=10$, in accordance to the pump-induced dephasing depicted in Fig.~\ref{f1}(c).

\begin{figure}
    \includegraphics[width=\linewidth]{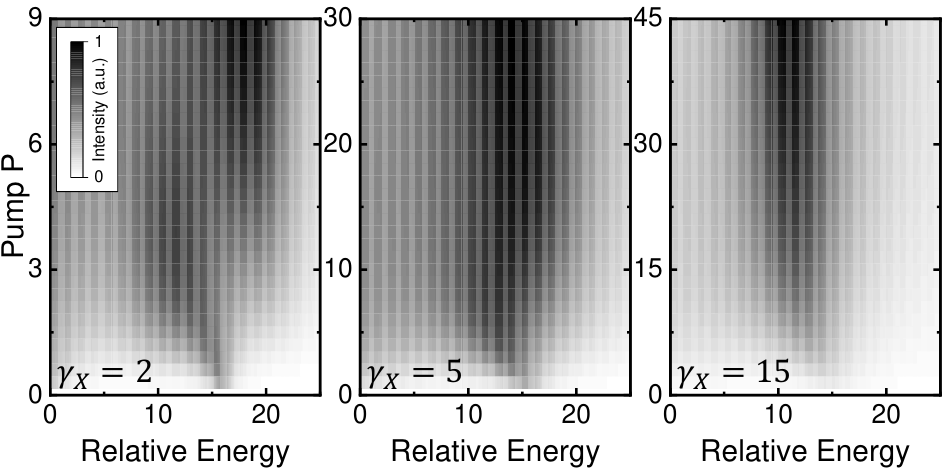}
    \caption{\label{f2}
        Calculated emission spectra around the bare cavity energy, with $N_{C(X)}=2$ and various emitter linewidth $\gamma_X$.
        With $\gamma_X=2$ and $5$, the pump of system to high order couplings is observed from the nonmonotonic peak energy shift.
        In contrast, with $\gamma_X=15$, the pump-induced dephasing dominates throughout, thus only monotonic shift is observed.
    }
\end{figure}

For the system with narrow-linewidth emitters ($\gamma_X=0.25$) presented in Fig.~\ref{f1}, the mesoscopic excitation level $P \in \left(0.4,\ 100\right)$ is clearly observed in the emission peaks.
Next, we investigate the dependence of this phenomenon on the emitter decay rate (linewidth) $\gamma_X$.
In Fig.~\ref{f2}, we present the emission spectra calculated with various $\gamma_X$.
Other parameters $\omega_X=0$, $\omega_C=10$, $\gamma_C=0.2$, $g_X=10$ are same to those used in Fig.~\ref{f1}.
The number of photons (emitters) is fixed at $N_{C(X)}=2$.
We focus on the spectral emission around the bare cavity energy $\omega_C=10$.
As shown in Fig.~\ref{f2}, for $\gamma_X=2$, two distinguishable peaks from the low (bottom) and high (top) order coupling are clearly observed, along with the converting of dominated emission peak at $P=5$.
For $\gamma_X=5$, although only one emission peak is distinguishable, the nonmonotonic energy shift clearly reveals that the coupling is pumped to high orders before suppressed by the pump-induced dephasing.
In contrast, for $\gamma_X=15$, only one emission peak with monotonic energy shift is observed, indicating the dephasing effect dominates throughout.
These results reveal that the nonmonotonic peak energy shift as the feature of high order couplings is only observed for $\gamma_X$ below a threshold (lifetime above a threshold).

The threshold value of $\gamma_X$ is further calculated as 10.0 for the case in Fig.~\ref{f2}.
This is much smaller compared to the threshold of Rabi splitting in the first-order coupling i.e., $4g_X+\gamma_C$ which has the value 40.2 here \cite{PhysRevB.60.13276,Khitrova2006}.
The calculation of threshold value in more cases are presented in Appendix~\ref{tel}.

\subsection{\label{sec2c}Ensemble of Emitters in Real Cavities}

\begin{figure}
    \includegraphics[width=\linewidth]{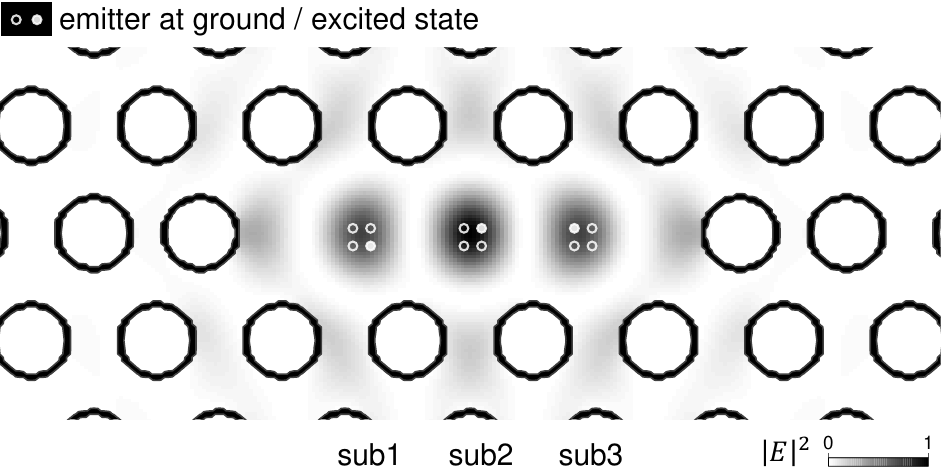}
    \caption{\label{f3}
        One example of typical L3 cavity.
        The cavity mode only couples to the emitters at the three antinodes, thereby, the emitters are subgrouped into three.
        In each subgroup, at most one excitation is allowed to exclude the emitter-emitter interaction.
        This is equal to three subcollective emitters, corresponding to the case $N_X=3$ in our model.
    }
\end{figure}

To bridge our theoretical calculation with the experiment, we start by discussing a cavity mode that couples to totally $N_{total}$ number of emitters, with the same coupling strength $g_j$ to each emitter.
As an example, Fig.~\ref{f3} shows a typical L3 photonic crystal cavity that couples to totally 12 emitters at the antinodes \cite{PhysRevB.60.13276,doi:10.1063/1.4961389}.
Usually, not all emitters can be at the excited state simultaneously \cite{PhysRevX.11.031033}, since the emitter-emitter interaction will be significant when two excited emitters are close to each other in spatial.
The emitter-emitter interaction will result in complex effects \cite{PhysRevB.103.045426,Kremser2020,Regan2022} thus dephase the coupling to the cavity \cite{PhysRevB.82.113106,Wu_2019}.
Therefore, the cavity equally couples to $N_X$ subcollective emitters.
Each subcollective emitter has $N_{total}/N_X$ emitters which are close to each other in spatial, and among the $N_{total}/N_X$ at most one excitation is allowed to exclude the emitter-emitter interaction.
E.g., in Fig.~\ref{f3} the 12 emitters are subgrouped into three subcollective emitters.
Thus, each subcollective emitter has a coupling strength to the cavity mode $g_i=\sqrt{N_{total}/N_X}g_j$ \cite{PhysRevA.105.013719,10.1109/TAC.2022.3169582}.
The first-order coupling strength between the cavity mode and $N_X$ subcollective emitters is $g_X=\sqrt{N_X}g_i=\sqrt{N_{total}}g_j$.
Meanwhile, the system also supports high energy states since the $N_X$ subcollective emitters can be simultaneously excited.
As shown, this case is exactly described by the calculation with $N_X$ emitters in Sec.~\ref{sec2b}.
Meanwhile, the limited number of photons in the calculation is allowed to exclude the dephasing effects such as the photon-phonon interaction \cite{RevModPhys.86.1391}.
Therefore, our model is a good approximation of the cavity QED system involving multiple excitations in experiment.

\section{\label{sec3}Experimental Verification}

\begin{figure}
    \includegraphics[width=\linewidth]{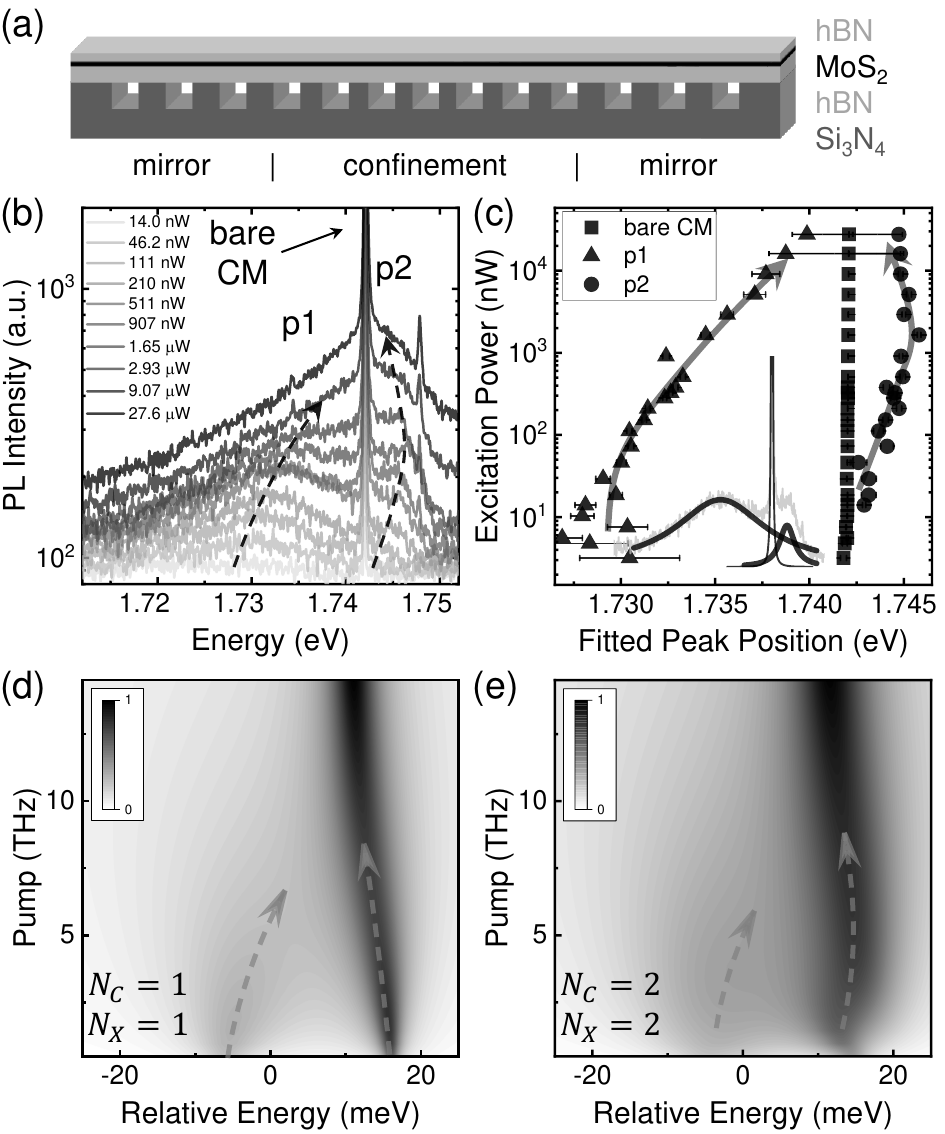}
    \caption{\label{f4}
        (a) Schematic of the cavity structure.
        (b) Power-dependent PL spectra recorded from the cavity-LXs system.
        Besides the bare cavity mode (CM) at 1.742 eV, mainly two emission peaks denoted by p1 and p2 are observed.
        The minor sharp peak at 1.748 eV is just noise from a neighbor cavity.
        (c) Fitted peak energies, along with one fitting example in the inset.
        (d)(e) Calculated emission spectra with $N_{C(X)}=1,2$, respectively.
        The nonmonotonic energy shift of p2 observed in (b)(c) is well explained by the calculation in (e).
    }
\end{figure}

The \textit{localized} excitons (LXs) in monolayer MoS$_2$ with emissions around 1.75 eV are a good example of \textit{individual} emitters, since their wavefunctions are strongly localized at defects \cite{Klein2019,Mitterreiter2021,Trainer2022}.
The emission energy of each LX is not exactly same, and the emission peak is asymmetric due to phonon sidebands \cite{Klein2019}.
Nonetheless, recent work shows that an inhomogeneously broadened ensemble of emitters also collectively couples to high-Q cavity mode \cite{PhysRevA.100.053821}.
We fabricate high-Q nanobeam cavities with monolayer MoS$_2$ embedded as schematically presented in Fig. \ref{f4}(a) and investigate the cavity-LXs coupling using PL spectroscopy.
The cavity is based on the hBN/MoS$_2$/hBN/Si$_3$N$_4$ heterostructure, and details of fabrication methods have been reported in previous literature \cite{PhysRevLett.128.237403,PhysRevLett.130.126901}.
The sample is excited by a 532-nm cw-laser with a spot size $\sim 1\ \mathrm{\mu m}$.
The power-dependent PL spectra recorded at low temperature (11 K) are presented in Fig.~\ref{f4}(b).
At this temperature, an individual localized exciton has the linewidth at the magnitude around 1 meV \cite{10.1021/acsphotonics.0c01907}.
The strong sharp peak at 1.742 eV with the linewidth of 0.16 meV arises from the bare cavity mode, corresponding to a Q-factor of $1.1 \times 10^4$.
This bare cavity peak is a widely observed phenomenon in cavity QED experiments, due to fluctuations of the system \cite{PhysRevLett.77.2901,Hennessy2007,Yamaguchi_2012} and background emissions which supply to the cavity mode \cite{10.1143/apex.2.122301,10.1063/1.5016615}.
The minor sharp peak at 1.748 eV is just noise arising from the bare cavity mode of a neighbor cavity.
That neighbor cavity is $1.5\ \mathrm{\mu m}$ ($>$ double photon wavelength) away in spatial, thus, has little impact here \cite{PhysRevLett.128.237403}.
Besides the two sharp peaks, we observe two polariton peaks as denoted by p1 and p2 in Fig.~\ref{f4}(b), arising from the cavity-LXs coupling.
Considering the typical defect density of 10$^{13}\ \mathrm{cm^{-2}}$ \cite{10.1038/srep29726} and the mode profile \cite{PhysRevLett.128.237403}, we estimate the number of LXs within the cavity mode volume to be $N_{total}=3\times10^5$.

Peak energies extracted by multi-Lorentz fittings are presented in Fig.~\ref{f4}(c), and one example of fitting is presented in the inset.
The fitting accuracy is limited by the factors that LXs are not homogeneously broadened, and their lineshape is not exactly Lorentzian \cite{Klein2019}.
In addition, at high excitation powers such as $27.6\ \mathrm{\mu W}$, the two polariton peaks merge to the bare cavity energy as predicted in Fig.~\ref{f1}(c), thus only one maximum of the broad polariton emission is observed.
For such broad polariton emission, the difference of peak energies between single and double peak fitting is included in the error bar in Fig.~\ref{f4}(c).
Nonetheless, the nonmonotonic energy shift of the right emission peak p2 is clearly observed both from the raw data in Fig.~\ref{f4}(a) and the fitted peak energies in Fig.~\ref{f4}(c).
In contrast, little power-dependent energy shift is observed in the control experiments recorded from MoS$_2$ outside cavities as presented in Appendix~\ref{cte}.
This distinct feature indicates that the cavity-LXs coupling is pumped to high orders as discussed in Sec.~\ref{sec2b}.

We use the parameters $\omega_X=0$, $\omega_C=10$, $\gamma_X=5$, $\gamma_C=0.2$, and $g_X=10$ to reproduce the cavity-LXs coupling, and typical results are plotted in Fig.~\ref{f4}(d)(e).
As expected, with $N_{C(X)}=1$ in Fig.~\ref{f4}(d), both two peaks shift monotonically.
In contrast, with $N_{C(X)}=2$ in Fig.~\ref{f4}(e), the right branch around the bare cavity energy exhibits the nonmonotonic energy shift as the pump increases, consistent with our experimental observation for p2.
We note that, the central result here is the high order couplings in the case of $N_{C(X)}>1$.
The specific value of $N_{C(X)}$ for the best fitting of experiment is not our focus, and is out of scope due to the limits in fitting accuracy.
In addition, the LXs in experiment are consistent to the ensemble of \textit{individual} emitters discussed in our theoretical calculations, due to they are \textit{localized} at defects in the monolayer MoS$_2$.
In contrast, \textit{free} excitons in semiconductors can transport in the material \cite{PhysRevLett.128.237403}, and even formulate the state of exciton gases with many-body interactions \cite{10.1002/adma.201706945,Jiang2021,Wilson2021,doi:10.1080/00018732.2021.1969727}.
Thus, the coupling between the cavity and \textit{free} excitons cannot be described by the model here.

\section{\label{sec4}Conclusion}

In summary, we investigate the cavity QED system with the high-$Q$ cavity mode and the ensemble of emitters using the master equation theory.
The increasing excitation of emitters has two effects: pump the system to high energy states and suppress the coherent coupling.
We observe the high order coupling between high energy states at mesoscopic excitation levels, when the emitter linewidth is smaller than the threshold.
Our model and calculation well describe the coupling between a high-$Q$ cavity and an ensemble of LXs in monolayer MoS$_2$, and is also expected to apply to other kinds of defect emitters such as defects in hBN, color centers in diamond, and rare-earth ions in silicon.
Such mesoscopic excitation levels have great potentials in the multi-photon based quantum information processing and nonlinear quantum photonic devices.

\begin{acknowledgments}

All authors gratefully acknowledge the German Science Foundation (DFG) for financial support via grants FI 947/8-1, DI 2013/5-1 and SPP-2244, as well as the clusters of excellence MCQST (EXS-2111) and e-conversion (EXS-2089).
C. Q. gratefully acknowledges support from the Chinese Academy of Sciences Project for Young Scientists in Basic Research (Grant No. YSBR-112) and the National Natural Science Foundation of China (Grants No. 12474426).
C. Q and V. V gratefully acknowledge the Alexander v. Humboldt foundation for financial support in the framework of their fellowship programme.

\end{acknowledgments}

\appendix

\section{\label{tel}Threshold of Emitter Linewidth}

\begin{figure}
    \includegraphics[width=\linewidth]{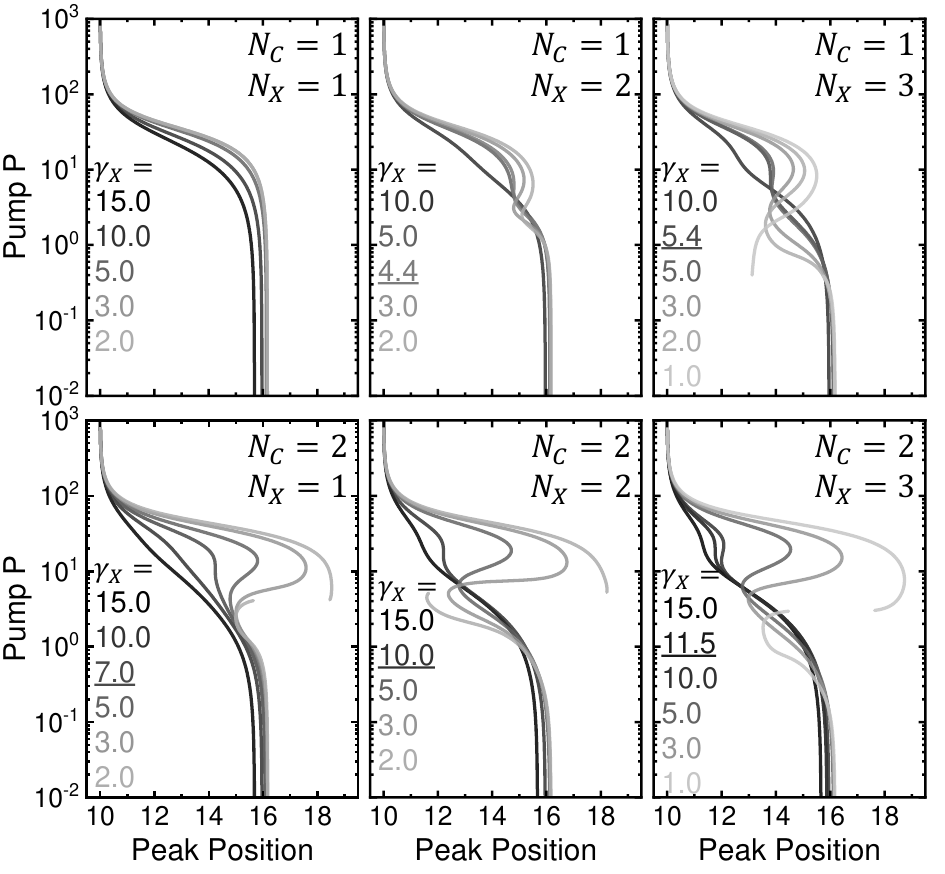}
    \caption{\label{f5}
        Energy shift of the emission peak around bare cavity energy, calculated in cases of various $N_C$ and $N_X$.
        The threshold value of $\gamma_X$ for nonmonotonicity is underlined.
    }
\end{figure}

The threshold of emitter linewidth is calculated in more cases with various $N_{C(X)}$, and the results are presented in Fig.~\ref{f5}.
With $N_{C(X)}=1$, as expected, the nonmonotonic shift never occurs, due to no high order coupling is allowed.
In other cases, the nonmonotonic shift occurs for emitter linewidth $\gamma_X$ smaller than a threshold as denoted in Fig.~\ref{f5}(a).
The threshold value increases with $N_{C(X)}$ due to the increasing high order coupling strengths \cite{PhysRev.170.379,PhysRev.188.692}.
In the calculation, $N_{C(X)}$ could further increase, and the nonmonotonic shift would become more significant.
However, the number of photons and emitters involved in the coupling cannot increase infinitely.
As $N_{C(X)}$ further increases, the system will become non-Markovian and the rotating wave approximation becomes invalid \cite{PhysRevLett.123.243602,PhysRevA.105.013719,PhysRevA.74.065801}.
The emitter-emitter interaction, photon-phonon coupling \cite{RevModPhys.86.1391}, experimental fluctuations \cite{PhysRevLett.77.2901,Hennessy2007,Yamaguchi_2012} and other complex effects will also dephase the coupling.
Moreover, as discussed in the context of Fig.~\ref{f4}, the specific value of $N_{C(X)}$ for the best fitting is not the scope of this work.
Therefore, we do not present the calculations with more $N_{C(X)}$ here.

\section{\label{cte}Control Experiments}

\begin{figure}
    \includegraphics[width=\linewidth]{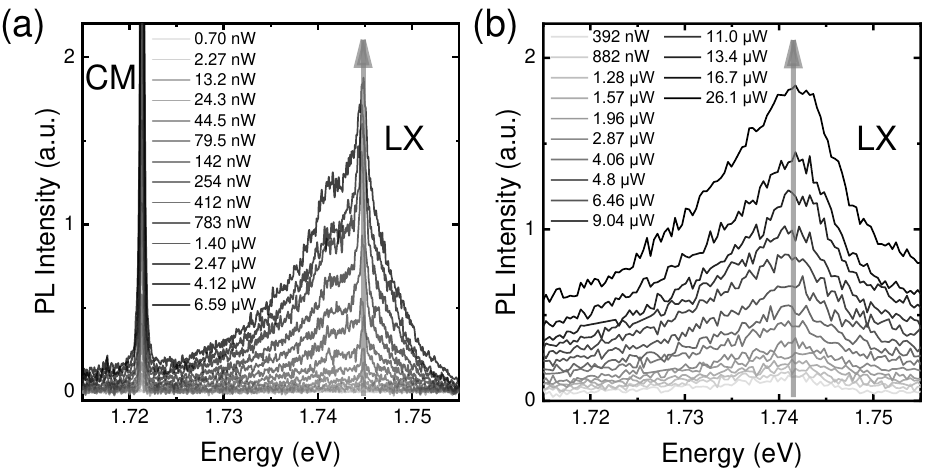}
    \caption{\label{f6}
        Experimentally recorded LX emissions (a) with the cavity mode (CM) far-away detuned and (b) when the MoS$_2$ is not embedded in the cavity.
        Little power-dependent energy shift of LX emissions is observed in the two cases.
    }
\end{figure}

For the experimental results in Fig.~\ref{f4}, one might wonder that, rather than the cavity-LXs coupling, the nonmonotonic shift arise from some intrinsic features of LX or the laser-induced heating \cite{10.1002/adma.200803616,PhysRevLett.120.037402}.
Hereby, we present the control experiments in Fig.~\ref{f6} to exclude these possibilities.
In Fig.~\ref{f6}(a) we present the PL spectra recorded in another cavity on the same sample, where the cavity mode at 1.721 eV is strongly detuned from the LX emissions.
In Fig.~\ref{f6}(b) we present the PL spectra recorded from MoS$_2$ on planar substrate without any nanostructures \cite{PhysRevLett.130.126901}.
In both cases, little power-dependent energy shift is observed when the excitation laser power is $<$ 30 $\mathrm{\mu W}$, as denoted by the gray arrow in Fig.~\ref{f6}.
Similar stable emission energy has also been reported for He ion generated LXs, even with the excitation laser power up to $500\ \mathrm{\mu W}$ \cite{Mitterreiter2021}.
These experiments further support that the nonmonotonic energy shift observed in Fig.~\ref{f4} arises from the high order cavity-LXs couplings.


%

\end{document}